\documentstyle[prl,aps,twocolumn]{revtex}
\begin{document}
\draft
\twocolumn[\hsize\textwidth\columnwidth\hsize\csname@twocolumnfalse%
\endcsname

\title{Spin transport in interacting quantum wires and carbon nanotubes}
\author{L.~Balents$^1$ and R.~Egger$^{2,3,\dagger}$}
\address{${}^1$~Physics Department, University of California, Santa
Barbara, CA 93106\\
${}^2$~Institute for Theoretical Physics, University of California,
Santa Barbara, CA 93106\\
${}^3$~Instituto Balseiro, Centro Atomico, 8400 S.C. de Bariloche,
Argentina
}
\date{Date: \today}
\maketitle

\begin{abstract}
  We present a general formulation of spin-dependent transport through
  a clean one-dimensional  interacting quantum wire or carbon
  nanotube, connected to non-collinear ferromagnets via tunnel
  junctions.  We show that the low energy description of each junction
  is given by a conformally-invariant boundary condition representing
  {\sl exchange coupling}, in addition to a pair of electron tunneling
  operators.   The effects of the exchange coupling are
  strongly enhanced by interactions, leading to a dramatic
  suppression of spin accumulation. This is a direct signature of
  spin-charge separation in a Luttinger liquid.  Furthermore, we
  demonstrate that magnetic polarization can lead to
  {\sl oscillations} in the non-linear current-voltage relation.  This
  phenomena is a surprising purely nonequilibrium effect due to
  backscattering interactions, which are thus dangerously marginally
  irrelevant in the repulsively-interacting Luttinger liquid.
\end{abstract}
]

\narrowtext

Recent studies on metal-ferromagnet hybrid systems have revealed new
and interesting physics due to the interplay between the electronic
charge and spin \cite{prinz}, e.g., the giant magnetoresistance effect
\cite{GMR}.  Following the initial spin-injection proposal
\cite{aronov}, the work of Johnson and Silsbee \cite{johnson} and
subsequent advances have opened the way for the field of spintronics,
where the electron spin is the central element for information storage
and transport\cite{Kikkawa}.  Spin-dependent transport plays an
important role in quantum computation proposals, and has already led
to new technological applications.  In this context, a detailed
understanding of transport through ferromagnetic-normal-ferromagnetic
devices is both of fundamental and technological interest.  In such
structures, the current-voltage relation is predicted to sensitively
depend on the relative angle $\theta$ between the magnetization
directions of the ferromagnets (FMs)\cite{GMR,slonc,brataas}.

Current theoretical models\cite{slonc,brataas}\ are based on Fermi
liquid theory, thereby ignoring the effect of interactions in the
metal.  As the inevitable miniaturization of spin-dependent devices
proceeds, however, at least the interconnects must ultimately reach
the one-dimensional (1D) quantum limit, in which Fermi liquid theory
breaks down\cite{book}.  This theoretically expected change from Fermi
liquid to {\sl Luttinger liquid} (LL) behavior drastically alters
transport phenomena, as has recently been verified in experiments on
charge conduction in carbon nanotubes\cite{LL-tubes2}, which are
nearly ideal 1D quantum wires (QWs)\cite{LL-tubes1}.  Despite these
developments, spin injection into a LL has received surprisingly
little attention \cite{si}.  In this paper, we present a general
low-energy theory for spin transport in a LL, which directly applies
to nanotubes and semiconductor QWs \cite{hammar}.  We assume, as
expected theoretically\cite{prelim}\ and recently observed
experimentally\cite{FM-tube}\ for carbon nanotubes, that spin-orbit
coupling in the LL is negligible. Its inclusion is, however,
straightforward.

Our analysis shows that spin transport in LLs is {\sl qualitatively}
different both from charge transport in LLs and from Fermi liquid spin
transport.  We focus for concrete results on the case of an
end-contacted quantum wire, and assume that the distance $L$ between
contacts is sufficiently long, ${\rm max} (V, T) \gg v/L$, to ensure
an incoherent stepwise transport mechanism through the tunnel barriers
between each FM and the QW.  (Here $v$ is the Fermi velocity, and we
put $e=k_B=\hbar=1$.)  A complete analytic solution of this problem is
contained in Eqs.~(\ref{eqm2}) and (\ref{tunnel}) to (\ref{curr}). 
In contrast to charge transport, we find that spin conduction 
occurs not only through
electron transfer but also exchange.  This exchange effectively gives
rise to a modification of the boundary conditions at the end of the
LL, e.g., for the left contact, 
\begin{equation} \label{bop} 
\vec{J}_R = {\cal R}(\Theta)\vec{J}_L + \vec{J}_{\rm tunnel} \;,
\end{equation}
where $\vec{J}_{L/R}$ is the left/right moving spin current into/out
of the contact, and $\vec{J}_{\rm tunnel}$ represents the effect of
electron transfer, see Eq.~(\ref{tunnel}).
The effect of exchange coupling is given by the
one-parameter $SO(3)$ matrix ${\cal R}(\Theta) = \exp(\Theta \Gamma)$, where
$\Gamma_{\mu\nu} = \sum_\lambda \hat{m}_\lambda
\epsilon_{\lambda\mu\nu}$, and $\hat{m}$ is a unit vector in the
direction of magnetization of the FM.  Physically, 
$\Theta$ represents the angle an incident spin in the LL
precesses due to exchange interaction with the FM.  Due to spin-charge
separation in the LL, the exchange contribution is {\sl not}\ suppressed by
the orthogonality catastrophe affecting the tunneling current, and
therefore dominates the physics in many situations.  This enhancement
of the exchange current does not occur in a Fermi liquid, and its
observation would provide a direct experimental signature of electron
fractionalization.  In addition to the novel physics arising at the
contact, we find that a long ballistic QW exhibits a {\sl bulk
precession} of the magnetization,
\begin{equation} \label{eqm2}
  v\partial_x\vec{M} + \partial_t  \vec{J} = b \vec{M}
  \times \vec{J}\;,
\end{equation}
where $\vec{M}=\vec{J}_R+\vec{J}_L$ and 
$\vec{J} = \vec{J}_R - \vec{J}_L$ are the local
magnetization and current in the QW, respectively.  Eq.~(\ref{eqm2})\ 
leads to {\sl oscillations} in the nonlinear current-voltage relation.
Remarkably, the latter is a purely non-equilibrium effect that arises
from a marginally irrelevant backscattering interaction
in the LL.  The detailed character of these oscillations is also
influenced by interactions.

We now turn to the derivation of these results.  In the incoherent
limit, we may consider each contact separately, as an initial system
composed of two decoupled pieces, $H_0= H_{FM}+H_{QW}$.  The FM
$(x<0)$, described by $H_{FM}$, is polarized along direction $\hat{m}$,
while the $SU(2)$ invariant QW
($x>0$) is described by $H_{QW}$.  The $SU(2)$ invariance guarantees
the existence of a continuity equation for spin density and current.
At time $t\to -\infty$, each half is assumed at equilibrium at its own
chemical potential, $\mu_{FM}$ and $\mu_{QW}$, and with a spin
chemical potential $\vec{h}$ (see below) in the QW.  We are interested 
in the steady state achieved at $t=0$, long after the tunneling
perturbation has been adiabatically turned on,
$H(t) = H_0 + e^{\delta t} H^\prime$  ($\delta \rightarrow 0^+$).
The calculation is
non-trivial primarily due to its non-equilibrium nature: the system
evolves according to $H(t)$ while the initial states are distributed
according to $\exp(-\beta H_0)$.  Consider
\begin{equation} \label{tunnelham}
  H' = F^\dagger W \Psi^{\vphantom\dagger} + \Psi^\dagger
  W^\dagger F^{\vphantom\dagger}\;, 
\end{equation}
where $F$ and $\Psi$ are spin-$1/2$ Fermion annihilation operators at
$x=0$ for the FM and the QW, respectively.  Employing the projection
operators $\hat{u}_s = (1\pm 
\hat{m}\cdot\vec\sigma)/2$, the $2 \times 2$ tunneling matrix $W$
takes the form $W = \sum_{s} t_s \hat{u}_s$, with spin-dependent
transmission coefficients $t_s$ for spin quantization axis parallel to
$\hat{m}$.  The junction is then characterized by the conductance
$G=G_\uparrow + G_\downarrow$ and the polarization
$P=(G_\uparrow-G_\downarrow)/G$, where
$G_{\uparrow,\downarrow}=(e^2/h)|t_{\uparrow,\downarrow}|^2$ are the
spin conductances \cite{brataas}.

$ $From Eq.~(\ref{tunnelham}) and the spin continuity equation, the
tunneling spin current across the junction is $\vec{J} =
-i\left(F^\dagger W \vec\sigma \Psi^{\vphantom\dagger} - \Psi^\dagger
  \vec\sigma \, W^\dagger F^{\vphantom\dagger} \right)/2$.  By defining
$\tilde{H}_0 = H_0 + \mu_{FM} N_{FM} + \mu_{QW} N_{QW} +\vec{h}\cdot
\vec{S}_{QW}$, the standard perturbative result can be rewritten as
\begin{eqnarray}
 \langle \vec{J} \, \rangle & = & {\rm Re}\,
\sum_{\alpha\beta\gamma\lambda} \int_{-\infty}^0 \! dt\,
 e^{\delta t}
 \left(W\vec\sigma\right)_{\alpha\beta}
 \left(U^\dagger(t)W^\dagger\right)_{\gamma\lambda}  \nonumber \\
 & & \times \left\langle
    \left[ F_\alpha^\dagger(0)\Psi_\beta^{\vphantom\dagger}(0),
      \Psi_\gamma^\dagger(t)F_\lambda^{\vphantom\dagger}(t)
      \right]  \right\rangle_{\tilde{H}_0}\;. \label{Jformula}
\end{eqnarray}
Thereby an intrinsically non-equilibrium expectation value is
expressed in terms of an equilibrium average using the shifted
Hamiltonian $\tilde{H}_0$, where the non-equilibrium nature of the
problem is fully encoded in the time dependent unitary matrix 
$U(t) = \exp[ i(V+\vec{h}\cdot\vec\sigma/2) t]$, 
with $V=\mu_{QW}-\mu_{FM}$.  A formula
similar to Eq.~(\ref{Jformula}) can easily be 
written down for the charge current, $I = i(F^\dagger W
\Psi^{\vphantom\dagger} - \Psi^\dagger W^\dagger
F^{\vphantom\dagger})$.  Thus both charge and spin current can be
calculated using equilibrium correlation functions.

To proceed, we specify the Hamiltonians $H_{FM}$ and $H_{QW}$.  For
energies well below the electronic bandwidth $D$, the $F$ and $\Psi$
{\sl equilibrium} correlators are identical for $H_0$ and
$\tilde{H}_0$, and moreover a non-interacting Fermi liquid model with
constant density of states (DOS) applies to the leads.
Because the lead couples to the QW only at $x=0$, the difference in
DOS for majority and minority spin carriers can be absorbed in a
spatial rescaling of the Fermi fields of the FM and a suitable
redefinition of the transmission coefficients ($t_s$)\cite{brataas}.
Then 
\begin{equation} \label{LFMa}
  H_{FM} = -i \int_{-\infty}^0 \!\!\!\!dx\, 
     f^\dagger \tau^z \partial_xf^{\vphantom\dagger}  \;,
\end{equation}
where the spinor $f = f_{b\beta}$ is indexed by
$b=(R,L)$, describing 
right- and left-moving modes of the FM, and by
$\beta=(\uparrow,\downarrow)$ for the spin, 
with the boundary condition $f_R(0)=f_L(0)$.
Here the Pauli matrix $\tau^z$ acts in the $R/L$ space. 
Putting $F = f(0)$, we see that $F$ has $SU(2)$ invariant correlation
functions.  The low-energy description of the QW is an
interacting LL model,
\begin{equation} \label{Lwire}
  H_{QW} = \int_0^\infty\!\!\!\!dx\, \left\{
    -i \psi^\dagger
      v \tau^z \partial_x\psi^{\vphantom\dagger} + u
    \left(\psi^\dagger\psi^{\vphantom\dagger}\right)^2 \right\} \;,
\end{equation}
where $\psi_R(0) = \psi_L(0)$ and $\Psi=\psi(0)$.  Only the
forward-scattering interaction $u$ is kept in Eq.~(\ref{Lwire}).
Alternatively, the exponent $\alpha>0$ for tunneling into the end of
the LL ($x=0$) serves to measure the interaction strength \cite{book}.

The $SU(2)$ invariance of Eqs.~(\ref{LFMa}) and (\ref{Lwire}) implies 
\[
  \left\langle
    \left[F_\alpha^\dagger(0)\Psi_\beta^{\vphantom\dagger}(0),
      \Psi_\gamma^\dagger(t)F_\lambda^{\vphantom\dagger}(t) 
      \right]\right\rangle_{\tilde{H}_0} \theta(-t) =
    \delta_{\alpha\lambda} \delta_{\beta\gamma} iC(-t) \;,
\]
where $C(t)$ is the retarded Green's function of the operator 
$F_\uparrow^\dagger \Psi_\uparrow$ (the choice 
of spin components is arbitrary).  From Eq.~(\ref{Jformula}) 
and the corresponding expression for $I$, 
it is then straightforward to obtain 
\begin{eqnarray}
  \langle \vec{J} \;\rangle & = & -{G \over 2}\sum_{s}
  \bigg[ (P\hat{m}+s\hat{h})\; {\rm Im}\; \tilde{C}(V+hs/2+i\delta)
  \nonumber \\
  & & - P s \hat{m}\times\hat{h}\;{\rm Re}\,
  \tilde{C}(V+hs/2+i\delta)\bigg]\;, \label{Jeq1} \\
  \left\langle I \right\rangle & = & - G\sum_{s}
  (1+P s\hat{m}\cdot\hat{h}) \;{\rm Im} \,
  \tilde{C}(V+hs/2+i\delta)\;, \label{Ieq1}
\end{eqnarray}
where $\tilde{C}(V) = \int\! dt \; C(t) e^{i Vt}$.
The terms involving  \cite{book}
\begin{eqnarray} \label{scaling}
&&{\cal I}_\alpha(V,T) \equiv G \,{\rm Im}\, \tilde{C}(V+i\delta) \\ \nonumber
&&= G T (T/D)^{\alpha} \sinh(V/2T) \left| 
\Gamma\left( 1 + {\alpha \over 2} + i {V\over
      {2\pi T}}\right)\right|^2
\end{eqnarray}
have a simple interpretation in terms of tunneling via Fermi's
golden rule, as can be seen from the spectral representation of
$\tilde{C}$.  
However, the appearance of ${\rm Re} \,\tilde{C}$ in
Eq.~(\ref{Jeq1}) indicates the presence of a physical process
other than tunneling. 
It can be shown \cite{slonc,prelim} that it corresponds to a
virtual process in which an electron near the Fermi energy in the QW
hops into a state of the FM (which could be far from the
Fermi energy) and hops back, thereby generating an {\sl exchange
coupling}.
We thus include it from the start, $H \rightarrow H_0 + H' +
H''$, with
\begin{equation} \label{K1}
  H'' = - K \hat{m}\cdot \Psi^\dagger \vec\sigma\,
  \Psi^{\vphantom\dagger}/2 \;.  
\end{equation}
Since the FM possesses a non-vanishing average magnetization,
the spin operator in the FM may be replaced by
this average to leading approximation.

It is helpful to view both $H'$ and $H''$ in
a renormalization group (RG) framework, as perturbations to a 
decoupled fixed point described by $H_0$.
Standard arguments give the scaling dimension of both $t_s$ and $K$,
$\Delta_{t_s} = 1+ \alpha/2$ and $\Delta_K = 1$.
The scaling dimension $\Delta_K$ is not renormalized due to
spin-charge separation in the QW.  A simple
calculation gives the RG scaling equations 
\[
  \partial_\ell|t_s|^2(\ell)  = -\alpha |t_s|^2 \;, 
\qquad \partial_\ell K(\ell)= c\left( |t_\uparrow|^2 -
    |t_\downarrow|^2\right) \;,
\]
where $\ell = \ln (D/E)$, and $c$ is a non-universal constant. 
Following the RG flow from the ultraviolet cutoff $D$ down to 
energy $E\approx {\rm max} \;(T,V)\ll D$, we find $|t_s|^2(E) = 
|t_s|^2 (E/D)^\alpha$, and
\[
K(E)  =  K + \alpha^{-1} c GP [1-(E/D)^\alpha] \gg |t_s|^2(E)\;.
\]
Therefore the tunneling spin current is much smaller than
the exchange contribution.  Neglecting the tunneling
contribution completely, one still obtains a
$T$-independent exchange spin current as $T\rightarrow 0$.

This fact can be understood from a simple analogy to the
Andreev current through a ballistic superconductor-normal-superconductor
(SNS) junction \cite{sns}. 
Let us consider a LL connected to two insulating FMs
at $x=0$ and $x=L$, with $\hat{m}\times\hat{m}' \neq 0$.
This is an equilibrium situation, which can be
modeled using Eq.~(\ref{Lwire}) for the LL and
two copies of Eq.~(\ref{K1}) for the contacts to the FMs. 
The exchange interaction operates entirely within the 
spin sector of the LL due to spin-charge separation. 
Since the charge sector is decoupled, we are free to 
consider it at the non-interacting point, $u=0$.  
Then the resulting fictitious charge boson and the physical spin boson
can be combined and refermionized into a spin-ful Dirac fermion
$\eta$.  
Choosing arbitrary quantization axes $\hat{m}=\hat{x}$ and 
$\hat{m}'=\cos(\theta)\hat{x}+\sin(\theta)\hat{y}$, it is instructive
to perform the particle-hole transformation $\eta_\downarrow \rightarrow
\eta_\downarrow^\dagger$.  This yields the Hamiltonian
\begin{equation} \label{Andreev}
 H_\eta  = -iv\int_0^L\!\!\!\! dx\; \eta^\dagger
   \tau^z \partial_x 
  \eta^{\vphantom\dagger}  -{\rm Re}\,[K \Delta(0) \!+\!
  K'\Delta(L)e^{i\theta} ] \;, 
\end{equation}
where $\Delta(x) = \eta_\uparrow(x)\eta_\downarrow(x)$.
Equation (\ref{Andreev}) describes a ballistic SNS junction,
and for phase twist $0<\theta<2\pi$ between the superconductors, 
supports an equilibrium current due to Andreev 
reflection.  Since the Andreev current
is $v \eta^\dagger \tau^z \eta^{\vphantom\dagger}$,
the original FM-LL-FM device indeed has a non-zero spin current $J_z$.
The analogy to a SNS junction also demonstrates that this current 
does not rely upon the incoherence of the two contacts.

A more general perspective on the exchange coupling can be gained by
viewing the low-energy physics entirely in terms of boundary operators
and boundary conditions \cite{cardy}.  For that purpose, we may make
an arbitrary choice of short-scale physics, and let the exchange
coupling act on right-movers slightly away from the junction.  In this
case, using the boundary condition $\psi_L(0)=\psi_R(0)$,
the equations of motion for the spin currents 
can be integrated over the junction region to give the steady-state relation
$\vec{J}_R(0^+) = {\cal R}(\Theta)\vec{J}_L(0^+)$
(the brackets denoting
expectation values are omitted henceforth).
The parameter $\Theta\approx K/v$ ultimately defines the
``exchange coupling constant'' of the low-energy theory.  In
principle, since the boundary exchange operator is exactly marginal,
$\Theta$ need not be small.  Then the ``bulk'' spin currents are
$\vec{J}_L = \vec{J}_L(0^+)$ and $\vec{J}_R = \vec{J}_R(0^+) + \vec{J}_{\rm
  tunnel}$, and we
obtain Eq.~(\ref{bop}), with the tunneling spin current
\begin{equation} \label{tunnel}
\vec{J}_{\rm tunnel}  = -{1 \over 2} \sum_s (P\hat{m}+s\hat{h}) \, {\cal
  I}_\alpha(V+hs/2, T) \;.
\end{equation}
The term proportional to ${\rm Re}\,\tilde{C}$ 
in Eq.~(\ref{Jeq1}) has been dropped, 
as its physical effects are included
via the $SO(3)$ rotation ${\cal R}$.  
Since the magnetization far from the
contact is $\vec{M}=\chi\vec{h}$ with
the LL spin susceptibility $\chi$, one can
then compute the spin current $\vec{J}$ for arbitrary
exchange coupling $\Theta$. 
We arrive at the general result \cite{foot}
\begin{equation}\label{spinc}
  \vec{J} =   S\chi \vec{h} + (1-S) \vec{J}_{\rm tunnel}\;,
\end{equation}
where $S=({\cal R}-1)/({\cal R}+1)$ is a real antisymmetric matrix.
Similarly, the charge current is
\begin{equation}\label{curr}
I  =  -\sum_s (1+s\hat{m}\cdot\hat{h})\,{\cal I}_\alpha(V+hs/2,T)\;.
\end{equation}
$ $From Eqs.~(\ref{spinc}) and (\ref{curr}), by exploiting
spin and charge current conservation in order to obtain $\mu_{QW}$
and $\vec{h}$, one can then 
compute all transport properties in a given circuit
for arbitrary parameters \cite{prelim}.  

We now specialize to a FM-LL-FM device
with identical contacts at $T=0$ and
applied voltage $V$ within $v/L\ll V\ll D$.
For algebraic simplicity, we require $P^2\ll (1+\alpha)^2$.
For a tunneling contact, one expects $\Theta \ll 1$ \cite{slonc}, 
and Eq.~(\ref{spinc}) then yields  
\[
 \vec{J} =  -(\Theta \chi/2) \,\hat{m} \times \vec{h}  + \vec{J}_{\rm tunnel}
  \;.
\]
Under these conditions, it is straightforward to obtain  
the $\theta$-dependent FM-LL-FM current-voltage relation, 
\begin{equation}\label{curr1}
I(\theta) = \frac{GV}{2} (V/D)^\alpha \left(1-P^2 
\frac{\tan^2(\theta/2)}{\tan^2(\theta/2) + Y_\alpha(V)} \right)\;,
\end{equation}
where $Y_\alpha(V) = 1 + (\Theta\chi/2)^2(1+\alpha)^{-2}
(V/D)^{-2\alpha}$.  For $\alpha\to 0$, this reproduces the result of
Ref.~\cite{brataas}.  Notably, unless the magnetizations of the FMs
are anti-parallel ($\theta=\pi$) or the exchange coupling vanishes
($\Theta=0$), the {\sl spin accumulation effect}\cite{prinz}, in which
the current is reduced due to pile-up of spin in the QW, is strongly
{\sl suppressed}\ by the voltage dependence of $Y_\alpha$.
Physically, this suppression is caused by the exchange coupling which
is much more efficient in relaxing the injected spin polarization
compared to the tunneling current.

Finally, we turn to backscattering electron-electron interactions of the form
\begin{equation} \label{pert}
H_b = -b \int_0^L\!\!\! dx\, \vec{J}_L \cdot \vec{J}_R \;.
\end{equation}
For a carbon nanotube, with the lattice spacing $a$ and the
tube radius $R$, one may estimate $b \approx a e^2/R$ \cite{LL-tubes1}. 
Since $H_b$ is marginally irrelevant
in a single-channel QW, it is usually
neglected in the LL model (\ref{Lwire}). 
Nevertheless, as is shown here,
{\sl dynamical}\ effects caused by (\ref{pert}) can
be important.  
The equations of motion away from the contacts give Eq.~(\ref{eqm2}) and
$v\partial_x\vec{J} +  \partial_t  \vec{M} = 0$.
In the steady state, we have conserved spin current $\vec{J}$,
and a {\sl bulk precession equation} for $\vec{M}$. 
Since $\vec{M} =\chi \vec{h}$, the vector $\hat{h}$ must
then precess around $\vec{J}$. 

To get sizeable consequences, detailed analysis \cite{prelim} shows
that it is essential to have small exchange couplings. For simplicity,
we put $\Theta=0$ below.  For the FM-LL-FM device,
\begin{equation} \label{phase}
\hat{h}(0)\cdot\hat{h}(L)=\cos(\Delta\varphi)\;,
\qquad \Delta \varphi = bJ L/ v \;.
\end{equation} 
Since $\Delta\varphi \propto L$, precession will then always be
significant for a sufficiently long QW.
The computation of $\vec{h}(0)$ and $\vec{h}(L)$ 
leads to the self-consistency equation
\begin{equation}
  (1-x^2) \cos\left(\frac{\pi x}{\cos(\theta/2)} \frac{V}{\Delta V}
    (V/D)^\alpha\right) = \sin^2(\theta/2) \;,
\end{equation}
where $\Delta V = 8\pi v/[GP b L \cos(\theta/2)]$.
Here solutions $x=x_n$ ($n\geq0$) in the
interval $0\leq x\leq \cos(\theta/2)$
have to be found.  The current through the device is then
\begin{equation}
I_n(V) = \frac{GV}{2} (V/D)^\alpha (1-P^2 [1-x_n^2(V)]) \;.
\end{equation}
The solution $I_n$ corresponding to $n$ full precession periods
exists only for voltages $V>V_n \equiv D [\Delta V/D]^{1/(1+\alpha)}
n^{1/(1+\alpha)}$.
Using typical parameters appropriate for a $1~\mu$m long single-wall
nanotube, one finds $V_1\approx 0.1$ to 1~V.  In general, the
current-voltage relation could then be multi-valued, where 
in the regime $V_n<V<V_{n+1}$, the solution $I_n(V)$ is expected
to be realized.
Hence the current-voltage relation becomes {\sl oscillatory},
with sawtooth-like oscillations.  The
observation of several periods could provide a direct and
accurate measurement of $\alpha$ via the $V_n$.  

To conclude, we have presented a general formalism for spin-dependent
transport through interacting 1D conductors.  An experimental check of
the theory should be possible by measuring the current-voltage
relation for a ferromagnet-nanotube-ferromagnet device.  The approach
is general enough to apply  to numerous other problems.  Several
interesting extensions currently under investigation are the
description of bulk-contacted wires, inclusion of the subband
degree of freedom of single-wall nanotubes, LL to LL contacts, and the
ballistic--diffusive crossover potentially 
relevant for multi-wall nanotubes.  

We thank Gerrit Bauer for helpful discussions.  Financial support was
provided by the NSF CAREER program under Grant NSF-DMR-9985255, 
the NSF grant PHY-94-07194, and by the DFG under the Gerhard-Hess and the
Heisenberg program.

{\small
$^{}\dagger$~Present address: 
Fakult{\"a}t f{\"u}r Physik, Universit{\"a}t Freiburg,
D-79104 Freiburg, Germany}


\begin{references} 


\bibitem{prinz} 
G.~A.~Prinz, Physics Today {\bf 48(4)}, 58 (1995).

\bibitem{GMR} 
For a review, see M.~A.~M.~Gijs and G.~E.~W.~Bauer,
Adv. Phys. {\bf 46}, 285 (1997), and references therein.


\bibitem{aronov} 
A.~G.~Aronov, JETP Lett. {\bf 24}, 32 (1976).

\bibitem{johnson} 
M.~Johnson and R.~H.~Silsbee, Phys. Rev. Lett.
{\bf 55}, 1790 (1985); 
M.~Johnson, {\sl ibid.} {\bf 70}, 2142 (1993).

\bibitem{Kikkawa} See, e.g., D.~D.~Awschalom and J.~M.~Kikkawa, Physics
  Today {\bf 52(6)}, 33 (1999), and references therein.

\bibitem{slonc} 
J.~Slonczewski, Phys. Rev. B {\bf 39}, 6995 (1989).

\bibitem{brataas} 
A.~Brataas {\sl et al.}, Phys. Rev. Lett. {\bf 84}, 2481 (2000).

\bibitem{book}
A.~O.~Gogolin, A.~A.~Nersesyan, and A.~M.~Tsvelik,
{\sl Bosonization and Strongly Correlated Systems}
(Cambridge University Press, 1998).


\bibitem{LL-tubes2} 
M.~Bockrath {\sl et al.}, Nature {\bf 397}, 598 (1999); 
Z.~Yao {\sl et al.}, {\em ibid.} {\bf 402}, 273 (1999).

\bibitem{LL-tubes1} 
R.~Egger and A.~O.~Gogolin, Phys. Rev. Lett.
{\bf 79}, 5082 (1997); C.~L.~Kane, L.~Balents, and M.~P.~A.~Fisher,
{\sl ibid.} {\bf 79}, 5086 (1997).

\bibitem{si} 
The only study that we are aware of 
deals with collinear magnetizations:
Q.~Si, Phys. Rev. Lett. {\bf 81}, 3191 (1998).

\bibitem{hammar}
Spin injection into semiconductors has
recently been achieved by 
P.~R.~Hammar {\sl et al.}, Phys. Rev. Lett. {\bf 83},
203 (1999); R.~Flederling {\sl et al.}, Nature {\bf 402}, 787 (1999);
Y.~Ohno {\sl et al.}, {\em ibid.} {\bf 402}, 790 (1999).

\bibitem{prelim}
L.~Balents and R.~Egger, (unpublished).

\bibitem{FM-tube}
K.~Tsukagoshi {\sl et al.}, Nature {\bf 401},
572 (1999).

\bibitem{sns}
A.~F.~Andreev, Sov. Phys. JETP {\bf 19}, 1228 (1964).

\bibitem{cardy}
J.~Cardy, {\sl Scaling and Renormalization in Statistical Physics}
(Cambridge University Press, Cambridge, 1996).

\bibitem{foot}
The (complex) mixing conductance $\eta$ of Ref.~\cite{brataas} 
is related to the exchange coupling.
For small $\Theta$, we find  $1+(\Theta\chi/2)^2 =  |\eta|^2/{\rm
Re}(\eta)$.

\end{references}
\end{document}